\documentclass[conference]{IEEEtran}
\IEEEoverridecommandlockouts
\usepackage[dvips]{graphicx}
\usepackage{algorithmic}
\usepackage[ruled,vlined,linesnumbered]{algorithm2e}
\usepackage{psfrag}

\usepackage{bm}
\usepackage{amssymb,mathtools}
\usepackage{amsfonts}
\usepackage{amsmath,amssymb}
\usepackage{amsthm}
\usepackage{cite} 
\usepackage[dvipsnames]{xcolor}
\usepackage{url}
\usepackage[letterpaper, left=0.625in, right=0.625in, bottom=1in, top=0.75in]{geometry}
\usepackage[hyperfootnotes=false,hidelinks]{hyperref}
\usepackage{upgreek}
\usepackage{yfonts}
\usepackage{cleveref}
\usepackage{cite}
\DeclareMathAlphabet\mathbfcal{OMS}{cmsy}{b}{n}
\renewcommand{\IEEEQED}{\IEEEQEDopen}
\usepackage{balance}
\usepackage[subrefformat=parens,labelformat=parens,caption=false]{subfig}
\usepackage{enumitem}
\SetKwComment{Comment}{}{}

\newcommand{\Ibf}{\mathbf{I}}

\newcommand{\Xbf}{\mathbf{X}}
\newcommand{\Ybf}{\mathbf{Y}}
\newcommand{\Zbf}{\mathbf{Z}}

\newcommand{\Acalbf}{\boldsymbol{\mathcal{A}}}

\newcommand{\Icalbf}{\boldsymbol{\mathcal{I}}}
\newcommand{\Dcalbf}{\boldsymbol{\mathcal{D}}}

\newcommand{\Rbb}{\mathbb{R}}

\DeclareMathOperator*{\st}{s.t.}

\DeclareMathAlphabet\mathbfcal{OMS}{cmsy}{b}{n}

\renewcommand{\IEEEQED}{\IEEEQEDopen}

\title{Privacy-Preserving Distributed Nonnegative Matrix Factorization}

\author{\IEEEauthorblockN{{Ehsan Lari}\textsuperscript{1}, \text{Reza Arablouei}\textsuperscript{2}, \text{Stefan Werner}\textsuperscript{1}}
\textsuperscript{1}\text{Department of Electronic Systems, Norwegian University of Science and Technology, Trondheim, Norway}
\\ \textsuperscript{2}\text{CSIRO's Data61, Pullenvale QLD 4069, Australia}
\thanks{This work was partially supported by the Research Council of Norway and the Research Council of Finland (Grant 354523).}
}
\makeatletter
\newcommand{\linebreakand}{%
  \end{@IEEEauthorhalign}
  \hfill\mbox{}\par
  \mbox{}\hfill\begin{@IEEEauthorhalign}
}

\begin{document}

\makeatother

\newcommand{\Lim}[1]{\raisebox{0.5ex}{\scalebox{0.8}{$\displaystyle \lim_{#1}\;$}}}

\maketitle

\begin{abstract}

Nonnegative matrix factorization (NMF) is an effective data representation tool with numerous applications in signal processing and machine learning. However, deploying NMF in a decentralized manner over ad-hoc networks introduces privacy concerns due to the conventional approach of sharing raw data among network agents. To address this, we propose a privacy-preserving algorithm for fully-distributed NMF that decomposes a distributed large data matrix into left and right matrix factors while safeguarding each agent's local data privacy. It facilitates collaborative estimation of the left matrix factor among agents and enables them to estimate their respective right factors without exposing raw data. To ensure data privacy, we secure information exchanges between neighboring agents utilizing the Paillier cryptosystem, a probabilistic asymmetric algorithm for public-key cryptography that allows computations on encrypted data without decryption. Simulation results conducted on synthetic and real-world datasets demonstrate the effectiveness of the proposed algorithm in achieving privacy-preserving distributed NMF over ad-hoc networks.

\end{abstract}

\section{Introduction} \label{intro}

Nonnegative matrix factorization (NMF) \cite{G14,C09,B07,6165290,NIPS2000_f9d11525} is a specific case of constrained low-rank matrix approximation \cite{U06} and a linear dimensionality reduction (LDR) technique aimed at representing nonnegative data more compactly through nonnegative factors. NMF, originally introduced as positive matrix factorization in \cite{P94}, has gained significant research interest, particularly after being popularized by \cite{L99}. It has found widespread applications in various fields such as signal and image processing, data mining and analytics, machine learning, and federated learning. Examples include air emission control \cite{P94}, visual object recognition \cite{S06}, video background-foreground separation \cite{K13}, spectral unmixing \cite{M14}, text mining \cite{GJ14}, blind source separation \cite{C08}, clustering \cite{T15}, collaborative filtering \cite{M10}, computational biology \cite{D08}, music analysis \cite{F09}, molecular pattern discovery \cite{B07}, efficient implementation of deep neural networks \cite{B16}, and detecting malware activities \cite{10179267}. Its popularity stems from its utility in identifying and extracting meaningful features from data in addition to serving as a powerful LDR technique.

Distributed optimization and estimation algorithms have garnered significant attention due to the ubiquity of data dispersed across multiple agents within network environments. The existing distributed NMF algorithms align with this trend, addressing scenarios where data is distributed among a network of agents. However, the conventional approach of sharing raw data among neighboring agents poses inherent security risks and compromises the privacy of sensitive information \cite{9063484,10032773,9217911}. Consequently, there is a pressing need for privacy-preserving distributed NMF algorithms that ensure the security and confidentiality of each agent's local data. 

The Paillier cryptosystem \cite{paillier1999public} is a fundamental tool for enhancing privacy in distributed algorithms. As a probabilistic asymmetric algorithm for public-key cryptography, it is specifically designed to provide secure homomorphic encryption. Its primary advantage lies in its homomorphic properties, which enable computations to be performed on encrypted data without decryption. The Paillier cryptosystem has proven to be effective in improving privacy in various applications, including smart grids \cite{9369360,6165271,7828093}, machine learning \cite{8653362}, smart homes \cite{8445571}, and federated learning \cite{9195012,10278720}. However, the potential advantages of its utilization in addressing the distributed NMF problem have not been explored in the literature.

In this paper, we introduce a privacy-preserving distributed NMF (PPDNMF) algorithm tailored for scenarios where the data matrix to be factorized is distributed among agents within an ad-hoc network. Each agent holds a subset of the columns of the data matrix. Our goal is to perform NMF of the entire data dispersed over the network in a fully distributed and secure manner. Specifically, agents participate in a distributed and collaborative process to estimate both the left and right factors, exchanging information exclusively with their immediate neighbors over secure communication links. While taking part in this collaborative process, agents maintain the privacy of their local data and the corresponding right factor estimates. We utilize the block coordinate-descent (BCD) algorithm and the alternating direction method of multipliers (ADMM) to develop our distributed NMF algorithm. Furthermore, to ensure privacy preservation, we integrate the Paillier cryptosystem into our algorithm. We evaluate the performance of the proposed algorithm through simulations using synthetic and real data, demonstrating its efficacy in achieving results comparable to those obtained by the centralized alternative.

\section{Distributed NMF}\label{sec:DNMF} 

The objective of NMF is to approximate a data matrix $\mathbf{Z} \in \Rbb^{L\times M}$ consisting of nonnegative entries using the product of left and right factor matrices, both with nonnegative entries. That is, $\mathbf{Z}=\mathbf{X}\mathbf{Y}$ where $\mathbf{X} \in \Rbb^{L\times K}$ and $\mathbf{Y} \in \Rbb^{K\times M}$, typically with $K\leq \min(L, M)$. This approximation represents the $L$-dimensional datapoints (columns of the data matrix) within a $K$-dimensional linear subspace spanned by the columns of the left factor, whose coordinates are given by the columns of the right factor. The nonnegativity constraint on the factors induces sparsity, further enhancing the compactness of the representation. Moreover, in many applications, the factors' nonnegativity is essential to their physical plausibility and intuitive interpretability. 

We utilize the least-squares criterion, which is appropriate when the perturbation in the data matrix $\mathbf{Z}$ can be modeled as a Gaussian process. Therefore, the NMF problem can be formulated as
\begin{align}\label{op0}
\min_{\mathbf{X},\mathbf{Y}} \hspace{1mm}& \frac{1}{2}\left\|\mathbf{Z}-\mathbf{X}\mathbf{Y}\right\|_\mathsf{F}^2\\
\st\hspace{1mm} & \mathbf{X}\geq0, \mathbf{Y}\geq 0,\notag
\end{align}
where $\|\cdot\|_\mathsf{F}$ denotes the Frobenius norm.
We consider the scenario where $\mathbf{Z}$ is distributed over a network with $N$ agents such that we have $\mathbf{Z}=\left[\mathbf{Z}_1,\cdots,\mathbf{Z}_N\right]$ and consequently $\mathbf{Y}=\left[\mathbf{Y}_1,\cdots,\mathbf{Y}_N\right]$ with $\mathbf{Y}_i \in \Rbb^{K\times M_i}$ and $\sum_{i=1}^{N} M_i = M$. Therefore, we rewrite \eqref{op0} as
\begin{align}\label{op1}
\displaystyle \min_{\mathbf{X},\left\{\mathbf{Y}_i\right\}} \hspace{1mm}& \frac{1}{2}\sum_{i=1}^{N}\left\|\mathbf{Z}_i-\mathbf{X}\mathbf{Y}_i\right\|_\mathsf{F}^2\\
\st \hspace{1mm}& \mathbf{X} \geq 0, \mathbf{Y}_i \geq 0.\notag
\end{align}
In a fully distributed approach, every agent, indexed by $i$, aims to estimate $\mathbf{X}$ and its own $\mathbf{Y}_i$ using its local data $\mathbf{Z}_i$ and by exchanging information solely with its immediate neighbors through single-hop communication.
To this end, we utilize the BCD algorithm and iteratively solve two optimization subproblems for $\mathbf{X}$ and $\mathbf{Y}$. That is, we repeat the following alternating minimizations until convergence is achieved:
\begin{align}
\mathbf{X}^{(n)}=\min_{\mathbf{X}} \hspace{1mm} & \frac{1}{2}\sum_{i=1}^N\left\|\mathbf{Z}_i-\mathbf{X}\mathbf{Y}_i^{(n-1)}\right\|_\mathsf{F}^2 \label{Xn} \\
\st \hspace{1mm} & \mathbf{X}\geq0, \notag \\
\mathbf{Y}_i^{(n)}=\displaystyle \min_{\mathbf{Y}_i} \hspace{1mm} & \frac{1}{2}\left\|\mathbf{Z}_i-\mathbf{X}^{(n)}\mathbf{Y}_i\right\|_\mathsf{F}^2, \forall i \in\{1,\dots,N\} \label{Yn} \\
\st \hspace{1mm} & \mathbf{Y}_i\geq0.  \notag
\end{align}
The superscript $(n)$ denotes the estimate of its respective parameter at the $n$th BCD iteration.

The solution of \eqref{Yn} can be localized straightforwardly, provided that each agent has access to the estimate $\mathbf{X}^{(n)}$.
To solve \eqref{Xn} in a fully distributed manner, we introduce the variable $\mathbf{X}_i$ at each agent $i$ as a local copy of $\mathbf{X}$ and enforce it to be equal to those of the agents within the immediate neighborhood of agent $i$, thereby achieving consensus across the network. Thus, we reformulate \eqref{Xn} into the following equivalent form
\begin{align}\label{Xni}
\mathbf{X}_i^{(n)}=\min_{\mathbf{X}_i} \hspace{1mm} & \cfrac{1}{2}\left\|\mathbf{Z}_i-\mathbf{X}_i\mathbf{Y}_i^{(n-1)}\right\|_\mathsf{F}^2+\imath(\mathbf{X}_i)\\\st \hspace{1mm} & \mathbf{X}_i=\mathbf{X}_j\ \ \forall j \in \mathcal{N}_i,\ \forall i\in\{1,\dots,N\} \notag
\end{align}
where $\imath(\cdot)$ denotes the indicator function accounting for the nonegativity constraint and $\mathcal{N}_i$ denotes the set of neighbors of agent $i$ with cardinality $d_i = |\mathcal{N}_i|$.
Subsequently, we decompose and decouple the optimization problems at the agents by introducing the auxiliary variables $\mathbf{U}_i,\mathbf{S}_{i,j}\in\mathbb{R}^{L \times K}$ and rewriting the optimization in \eqref{Xni} as
\begin{align}\label{q}
\min_{\mathbf{X}_i,\mathbf{U}_i,\mathbf{S}_{i,j}} \hspace{3mm} & \frac{1}{2}\left\|\mathbf{Z}_i-\mathbf{U}_i\mathbf{Y}_i^{(n-1)}\right\|_\mathsf{F}^2+\imath(\mathbf{X}_i) \\
& \mathbf{U}_i=\mathbf{X}_i \notag  \\
\st \hspace{6mm} & \mathbf{S}_{i,j}=\mathbf{U}_i\ \ \forall j \in \mathcal{N}_i,\ \forall i\in\{1,\dots,N\} \notag \\
& \mathbf{S}_{j,i}=\mathbf{S}_{i,j}. \notag 
\end{align}

We can express the corresponding aggregate augmented Lagrangian function as
\begin{align}\label{lagr}
\mathcal{L}&\left(\{\mathbf{X}_i\},\{\mathbf{U}_i\},\{\mathbf{S}_{i,j}\},\{\mathbf{P}_i\},\{\mathbf{Q}_{i,j}\}\right)  \notag \\
&=\cfrac{1}{2}\sum_{i=1}^{N}\left\|\mathbf{Z}_i-\mathbf{U}_i\mathbf{Y}_i^{(n-1)}\right\|_\mathsf{F}^2+\sum_{i=1}^{N}\imath(\mathbf{X}_i)\notag\\
&+\frac{\mu}{2}\sum_{i=1}^{N}\left\|\mathbf{X}_i-\mathbf{U}_{i}-\mathbf{P}_{i}\right\|_\mathsf{F}^2\notag\\
&+\frac{1}{2}\sum_{i=1}^{N}\sum_{j\in\mathcal{N}_i}\rho_{i,j}\left\|\mathbf{U}_i-\mathbf{S}_{i,j}-\mathbf{Q}_{i,j}\right\|_\mathsf{F}^2,
\end{align}
where $\mu$ and $\rho_{i,j}$ are penalty parameters and $\mathbf{P}_{i}, \mathbf{Q}_{i,j}\in\mathbb{R}^{L \times K}$ are scaled Lagrange multipliers. We maintain $\mu$ consistent across all agents and iterations. However, we allow each $\rho_{i,j}$, unique to the edge connecting agents $i$ and $j$, to vary over iterations~\cite{Z18}. Additionally, we consider $\rho_{i,j}=\rho_{j,i}\ \forall i,j$.

\subsection{Estimating the Left Factor} \label{XADMM}

Minimizing \eqref{lagr} using ADMM, which leads to the elimination of the auxiliary variables $\{\mathbf{S}_{i,j}\}$, yields the following iterations at each agent $i$~\cite{giannakis2017decentralized}:
\begin{align}
\mathbf{X}_i^{(n,m)}&=\Pi_{\ge 0}\left\{\mathbf{U}_i^{(m-1)}+\mathbf{P}_i^{(m-1)}\right\} \label{Xi} \\
\mathbf{U}_i^{(m)}&= \bigg[\mathbf{Z}_i\mathbf{Y}_i^{(n-1) \intercal} + \mu \left( \mathbf{X}_i^{(n,m)}-\mathbf{P}_i^{(m-1)} \right) \notag \\
& + \rho_i^{(m)} \left(\mathbf{U}_i^{(m-1)} + 2\mathbf{Q}_{i}^{(m-1)}-\mathbf{Q}_{i}^{(m-2)} \right) \bigg] \notag \\
& \times\left[\mathbf{Y}_i^{(n-1)}\mathbf{Y}_i^{(n-1)\intercal}+ \left(\mu+\rho_i^{(m)}\right)\mathbf{I}\right]^{-1} \label{Ui} \\
\mathbf{P}_i^{(m)}&=\mathbf{P}_i^{(m-1)}-\left(\mathbf{X}_i^{(n,m)}-\mathbf{U}_i^{(m)}\right) \label{Pi} \\
\mathbf{Q}_{i}^{(m)}&=\mathbf{Q}_{i}^{(m-1)}+ \sum_{j\in\mathcal{N}_i} \rho_{i,j}^{(m)} \left(\mathbf{U}_{j}^{(m)}-\mathbf{U}_{i}^{(m)}\right). \label{Qi}
\end{align}
Here, $(m)$ denotes the ADMM iteration index, $\Pi_{\ge 0}$ represents the projection onto the nonnegative orthant, and $(\cdot)^\intercal$ stands for matrix transpose. In addition, we define $\rho_i^{(m)}=\sum_{j\in\mathcal{N}_i} \rho_{i,j}^{(m)}$ and $\mathbf{Q}_i^{(m)} = \sum_{j\in\mathcal{N}_i} \rho_{i,j}^{(m)} \mathbf{Q}_{i,j}^{(m)}$. These ADMM iterations can be executed in a fully distributed manner, relying solely on locally available information and single-hop communications. Upon convergences of the algorithm, we utilize the latest estimates $\mathbf{X}_i^{(n,m)}$ for optimizing $\mathbf{Y}_i$ in the subsequent BCD iteration, i.e., $\mathbf{X}_i^{(n)}\leftarrow \mathbf{X}_i^{(n,m)}$.
Note that we enforce the nonnegativity constraint and consensus simultaneously.

\subsection{Estimating the Right Factor} \label{YADMM}

Similarly, we can employ ADMM to iteratively solve \eqref{Yn} as follows:
\begin{align}
\mathbf{Y}_i^{(n,k)}&=\Pi_{\ge 0}\left\{\mathbf{V}_i^{(k-1)}+\mathbf{R}_i^{(k-1)}\right\} \label{Yi} \\
\mathbf{V}_i^{(k)}&=\left(\mathbf{X}_i^{(n)\intercal}\mathbf{X}_i^{(n)}+\eta\mathbf{I}\right)^{-1} \notag \\
&\times\left[\mathbf{X}_i^{(n)\intercal}\mathbf{Z}_i+\eta\left(\mathbf{Y}_i^{(n,k)}-\mathbf{R}_i^{(k-1)}\right)\right] \label{Vi} \\
\mathbf{R}_i^{(k)}&=\mathbf{R}_i^{(k-1)}-\left(\mathbf{Y}_i^{(n,k)}-\mathbf{V}_i^{(k)}\right). \label{Ri}
\end{align}
Here, $(k)$ represents the ADMM iteration index and $\eta$ is the penalty parameter.
Once convergence is attained, we utilize the latest estimates $\mathbf{Y}_i^{(n,k)}$ to update $\mathbf{X}_i$ estimates in the subsequent BCD iteration, i.e., $\mathbf{Y}_i^{(n)}\leftarrow \mathbf{Y}_i^{(n,k)}$.

Note that, we employ warm start in both ADMM algorithms for estimating the left and right factors. At the onset of each BCD iteration, we initialize both ADMM inner iterations using the most recent estimates from the preceding iterations.

\subsection{Synchronization and Stopping} \label{StopCri}

Our algorithm does not require waiting for all agents to converge in either BCD or ADMM iterations. During $\mathbf{X}_i$ updates, the first agent to converge or reach a predetermined maximum number of ADMM iterations stops updating and raises a flag, signaling its neighbors to stop as well. This message is then propagated through the network until all agents stop updating. After completing the collaborative $\mathbf{X}_i$ update iterations, each agent can immediately start updating its $\mathbf{Y}_i$ estimate independently. When an agent stops $\mathbf{Y}_i$ update due to convergence or reaching the corresponding iteration limit, it begins the first iteration of $\mathbf{X}_i$ update and shares its $\mathbf{U}_i$ with its neighbors. Using warm start, in the first iteration of $\mathbf{X}_i$ update, each agent utilizes the neighbor $\mathbf{U}_j$ values from the previous BCD iteration. Afterwards, if an agent has not received all required $\mathbf{U}_j$ updates from its neighbors, subsequent iterations are postponed until they are all acquired, ensuring synchronization of $\mathbf{X}_i$ updates among all agents. To decide when to terminate the BCD iterations, one can employ the same strategy as described for $\mathbf{X}_i$ updates. That is, the initial agent to discern convergence or reach a predefined maximum number of BCD iterations halts its updates and notifies its neighbors. This notification propagates across the network until all agents cease updating.

\subsection{Convergence Analysis} \label{ConvPro}

We can express the optimization problem \eqref{q} as
\begin{align}
\min_{\boldsymbol{\mathcal{X}},\boldsymbol{\mathcal{U}}}\ & f\left(\boldsymbol{\mathcal{X}}\right)+g\left(\boldsymbol{\mathcal{U}}\right)\\
\st\ & \mathbf{A}\boldsymbol{\mathcal{X}}+\mathbf{B}\boldsymbol{\mathcal{U}}=\mathbf{0},
\end{align}
where
\begin{align}
&f\left(\boldsymbol{\mathcal{X}}\right)=\sum_{i=1}^{N}\imath(\mathbf{X}_i), \
g\left(\boldsymbol{\mathcal{U}}\right)=\cfrac{1}{2}\sum_{i=1}^{N}\left\|\mathbf{Z}_i-\mathbf{U}_i\mathbf{Y}_i^{(n-1)}\right\|_\mathsf{F}^2, \notag \\
&\boldsymbol{\mathcal{X}}=\begin{bmatrix}\mathbf{X}_1,\cdots,\mathbf{X}_N\end{bmatrix}^{\intercal}\hspace{-2pt}, \hspace{2pt}\boldsymbol{\mathcal{U}}=\begin{bmatrix}\mathbf{U}_1,\cdots,\mathbf{U}_N\hspace{2pt} \vline\hspace{2pt} \mathbf{S}_{1,1},\cdots,\mathbf{S}_{N,N}\end{bmatrix}^{\intercal}\hspace{-2pt}, \notag
\end{align}
\begin{equation}
\mathbf{A}=\left[
\begin{array}{c}\mathbf{I}\\\hline\mathbf{0}\\\hline\mathbf{0}
\end{array}\right],\ \mathbf{B}=\left[\begin{array}{c|c}-\mathbf{I}&\mathbf{0}\\\hline \Dcalbf & -\Acalbf  \\\hline \mathbf{0} & \Icalbf\end{array} \right], \notag
\end{equation}
and $\Dcalbf = \mathrm{bdiag} (d_1 \Ibf_{L\times K},\cdots,d_N \Ibf_{L\times K})$. In addition, $\Acalbf \in \Rbb^{NL \times N^2K}$ and $\Icalbf \in \Rbb^{NL \times N^2K}$ represent modified versions of the adjacency and edge-node incidence matrices of the network, respectively \cite{6425904}. Consequently, the convergence of the ADMM iterations \eqref{Xi}-\eqref{Qi} can be proven using the approach proposed in \cite{deng2016global,wang2019global,boyd2011distributed}. The convergence of \eqref{Yi}-\eqref{Ri} can also be verified in a similar manner.

\section{Privacy-Preserving Distributed NMF}

\begin{algorithm}[t!]\label{Alg1}
\renewcommand{\AlCapSty}[1]{\normalfont{#1}\unskip}
\caption{The PPDNMF algorithm as agent $i$.}
\linespread{1.4}\selectfont
\scriptsize
\DontPrintSemicolon
\SetNlSty{text}{}{}
$\mathbf{X}_i^{(0)}=\mathbf{U}_i^{(0)}=\mathbf{1}_{L\times K}$,
$\mathbf{P}_i^{(0)}=\mathbf{Q}_i^{(0)}=\mathbf{Q}_i^{(-1)}=\mathbf{0}_{L\times K}$,\\
$\mathbf{Y}_i^{(0)}=\mathbf{V}_i^{(0)}=\mathbf{R}_i^{(0)}=\mathbf{0}_{K\times M_i}$\\
\For{$n=1,2,\hdots,$ until convergence}{
\For{$m=1,2,\hdots,$ until convergence}{
$\mathbf{X}_i^{(n,m)}=\Pi_{\ge 0}\left\{\mathbf{U}_i^{(m-1)}+\mathbf{P}_i^{(m-1)}\right\}$\;
$\mathbf{U}_i^{(m)}=\left[\mathbf{Z}_i\mathbf{Y}_i^{(n-1)\intercal}+\mu\left(\mathbf{X}_i^{(n,m)}-\mathbf{P}_i^{(m-1)}\right)\right.$\;
$\hspace{7.5mm}\left.+\rho_{i}^{(m)}\left(\mathbf{U}_i^{(m-1)} + 2\mathbf{Q}_{i}^{(m-1)}-\mathbf{Q}_{i}^{(m-2)} \right)\right]$\;
$\hspace{7.5mm}\times\left[\mathbf{Y}_i^{(n-1)}\mathbf{Y}_i^{(n-1)\intercal}+\left(\mu+\rho_{i}^{(m)}\right)\mathbf{I}\right]^{-1}$\;
$\mathbf{P}_i^{(m)}=\mathbf{P}_i^{(m-1)}-\left(\mathbf{X}_i^{(n,m)}-\mathbf{U}_i^{(m)}\right)$\;
encrypt $-\mathbf{U}_{i}^{(m)}$ using the public key $k_{pi}$ as $\mathcal{E}_i \left(-\mathbf{U}_{i}^{(m)}\right)$\;
send $\mathcal{E}_i\left(-\mathbf{U}_{i}^{(m)}\right)$ and $k_{pi}$ to all neighbors in $\mathcal{N}_i$

\For{$j \in \mathcal{N}_i$}{
encrypt $\mathbf{U}_{j}^{(m)}$ using $k_{pi}$ as $\mathcal{E}_i\left(\mathbf{U}_{j}^{(m)}\right)$



send $\left[\mathcal{E}_i\left(\mathbf{U}_{j}^{(m)}\right)\mathcal{E}_i\left(-\mathbf{U}_{i}^{(m)}\right)\right]^{{g}_{j\to i}^{(m)}}$ to agent $i$
}
decrypt messages received from neighbors using the private key $k_{si}$ and multiply them by $ {g}_{i \to j}^{(m)}$\;
$\mathbf{Q}_{i}^{(m)}=\mathbf{Q}_{i}^{(m-1)} + \sum_{j\in\mathcal{N}_i} {g}_{i \to j}^{(m)} {g}_{j \to i}^{(m)}\left(\mathbf{U}_{j}^{(m)} - \mathbf{U}_{i}^{(m)}\right)$.\;
}
$\mathbf{X}_i^{(n)}\leftarrow \mathbf{X}_i^{(n,m)}$,
$\mathbf{U}_i^{(0)}\leftarrow \mathbf{U}_i^{(m)}$,
$\mathbf{P}_i^{(0)}\leftarrow \mathbf{P}_i^{(m)}$,\;
$\mathbf{Q}_i^{(0)}\leftarrow \mathbf{Q}_i^{(m)}$,
$\mathbf{Q}_i^{(-1)}\leftarrow \mathbf{Q}_i^{(m-1)}$\;
\For{$k=1,2,\hdots,$ until convergence}{
$\mathbf{Y}_i^{(n,k)}=\Pi_{\ge 0}\left\{\mathbf{V}_i^{(k-1)}+\mathbf{R}_i^{(k-1)}\right\}$\;
$\mathbf{V}_i^{(k)}=\left(\mathbf{X}_i^{(n)\intercal}\mathbf{X}_i^{(n)}+\eta\mathbf{I}\right)^{-1}$\;
$\hspace{7.2mm}\times\left[\mathbf{X}_i^{(n)\intercal}\mathbf{Z}_i+\eta\left(\mathbf{Y}_i^{(n,m)}-\mathbf{R}_i^{(k-1)}\right)\right]$\;
$\mathbf{R}_i^{(k)}=\mathbf{R}_i^{(k-1)}-\left(\mathbf{Y}_i^{(n,k)}-\mathbf{V}_i^{(k)}\right)$\;}
$\mathbf{Y}_i^{(n)}\leftarrow \mathbf{Y}_i^{(n,k)}$,
$\mathbf{V}_i^{(0)}\leftarrow \mathbf{V}_i^{(k)}$,
$\mathbf{R}_i^{(0)}\leftarrow \mathbf{R}_i^{(k)}$\;}
\end{algorithm}

\begin{figure*}[t!]
\centering
\subfloat[\label{fig:1}]{{\includegraphics[width=.333\textwidth, height = 50mm]{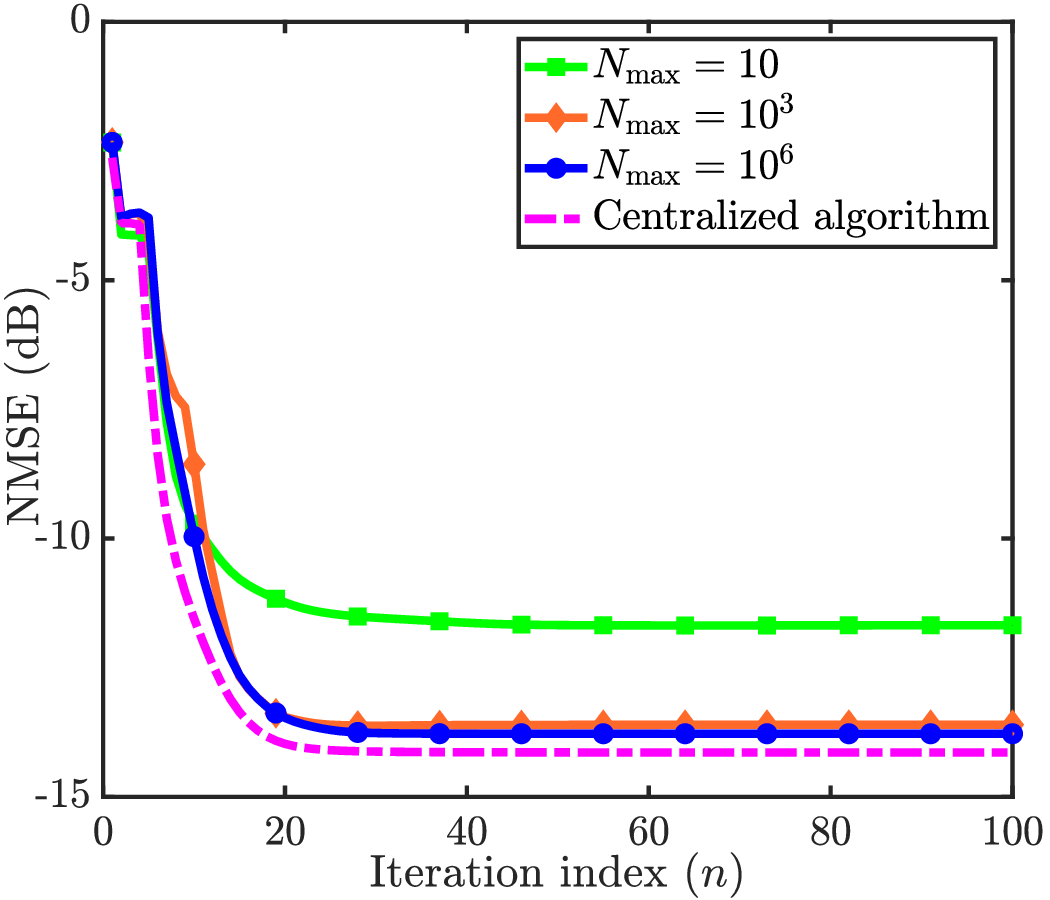}}}
\subfloat[\label{fig:2}]{{\includegraphics[width=.333\textwidth, height = 50mm]{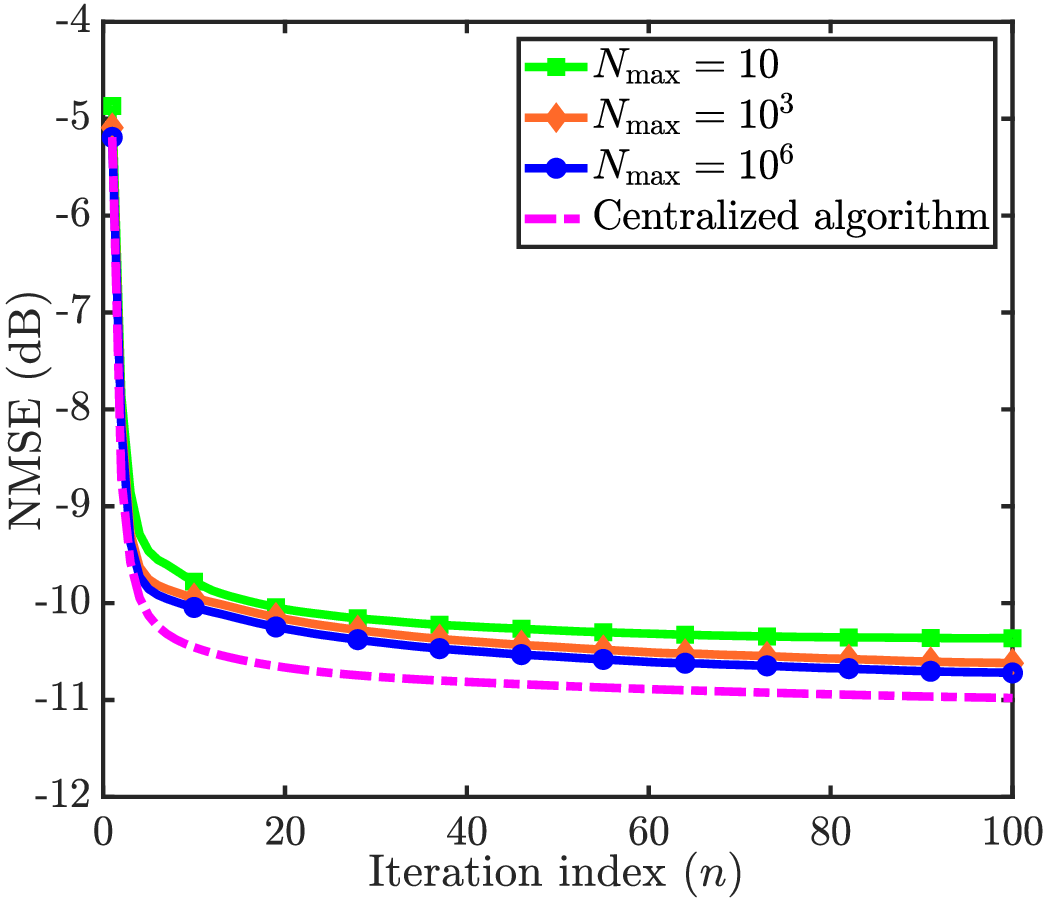}}}
\subfloat[\label{fig:3}]{{\includegraphics[height =50mm]{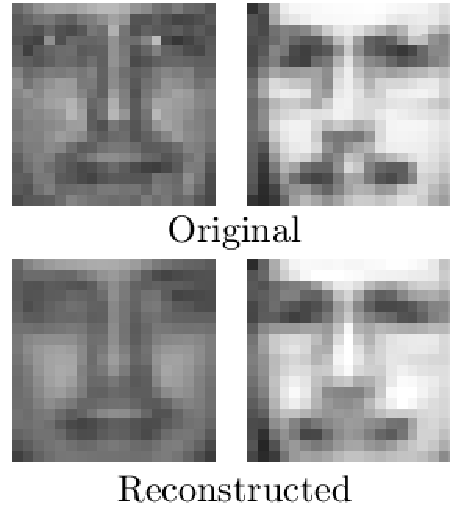}}}
\caption{The normalized mean-square error (NMSE) values of PPDNMF and centralized algorithm versus the BCD iteration index for different values of $N_{\max}$ on (a) synthetic data and (b) MIT-CBCL database. (c) The original and reconstructed faces \#1 and \#2429 from the MIT-CBCL database.}
\label{fig:EUSIPCO}
\end{figure*}

In this section, we provide a brief overview of the Paillier cryptosystem, which we employ to enhance the privacy of the distributed NMF algorithm developed in section \ref{sec:DNMF}. Subsequently, we introduce our proposed privacy-preserving distributed NMF (PPDNMF) algorithm.

\subsection{Paillier Cryptosystem}

In the Paillier cryptosystem, a public key and a private key are utilized. The public key is broadcast publicly, allowing other users to encrypt messages. However, decrypting the messages is only possible with the private key, which remains unknown to other users. This cryptosystem exhibits additive homomorphism \cite{7403296,8601358,Z18}, i.e., $\mathcal{E}\left( m_3 \left(m_1+m_2\right) \right) = \left( \mathcal{E}\left(m_1\right)
\mathcal{E}\left(m_2\right)\right)^{m_3}$.

\subsection{Privacy Preservation}

The only update equation among~\eqref{Xi}-\eqref{Ri} that relies on information received from neighboring agents is~\eqref{Qi}. To protect the privacy of agents at this step, we adopt a similar approach to~\cite{Z18} and enable the agents to encrypt all messages communicated with their neighbors. To this end, we decompose each edge-specific penalty parameter as $\rho_{i,j}^{(m)} = {g}_{i \to j}^{(m)}{g}_{j \to i}^{(m)}$ where ${g}_{i \to j}^{(m)}$ and $ {g}_{j \to i}^{(m)}$ are exclusively known to agents $i$ and $j$, respectively. In addition, we implement a secure data exchange procedure as outlined in lines 10-15 of Algorithm~\ref{Alg1}, which provides a summary of the proposed PPDNMF algorithm. Consequently, note the following:

\begin{itemize}[leftmargin=*]
\item Data exchanged between agents $i$ and $j$ is encrypted, rendering it inaccessible to other agents or eavesdroppers, even if intercepted.
\item The parameter ${g}_{i \to j}^{(m)}$ is unique to each edge and iteration. Therefore, an agent cannot infer the private information $\mathbf{U}_{j}^{(m)}$ of any of its neighbors by decrypting the messages it receives from them, as each neighbor $j$ uses its unique ${g}_{j \to i}^{(m)}$ in its encrypted message to agent $i$. 
\item The Paillier cryptosystem is intended for encrypting scalar unsigned integers. To encrypt the entries of $\mathbf{U}_{i}^{(m)}$, which are typically floating-point values, we initially quantize them. This involves multiplying each entry by a positive integer $N_{\max}$, which determines the quantization resolution, and then rounding the result to the nearest integer. To undo the quantization, we divide the decrypted values by $N_{\max}$.
\item To guarantee convergence, we ensure that the parameters ${g}_{i \to j}^{(m)}$ increase monotonically over iterations without becoming unbounded~\cite{Z18}. Thus, we select each parameter uniformly from the interval $\left({g}_{i \to j}^{(m-1)},g_i\right]$, where $g_i$ is a predefined positive constant, known only to agent $i$, and ${g}_{i \to j}^{(0)}=0$.
 
\end{itemize}

\section{Simulation Results}

In this section, we conduct a series of numerical experiments to evaluate the performance of our PPDNMF algorithm. We consider a network consisting of $N = 10$ agents, interconnected arbitrarily, with each agent having three neighbors on average. We test our algorithm on two datasets, namely, a synthetic dataset and the MIT-CBCL face database~\cite{fischercbcl}. 

The agents collaboratively factorize a data matrix $\Zbf \in \Rbb^{L \times M}$ to left and right factor matrices $\Xbf \in \Rbb^{L \times K}$ and $\Ybf \in \Rbb^{K \times M}$, where $L \in \{30,361\}, M \in \{200,2429\}$, and $K \in \{ 5,49 \}$ in our two experiments. We set the number of BCD iterations to $100$ and the number of ADMM iterations to $30$. In our implementation of the Paillier cryptosystem, we use $128$-bit public and private keys. To handle the encryption of negative quantized values (note line 10 in Algorithm \ref{Alg1}), we convert them to positive integers by adding the public key to them~\cite{7403296}.

We evaluate the performance of PPDNMF in comparison with the centralized algorithm, i.e., where all data is available at a central hub. To quantify the performance, we utilize the normalized mean-square error (NMSE) at each BCD iteration, defined as $\frac{1}{N} \sum_{i=1}^{N} \| \Zbf_i - \Xbf_{i}^{(n)} \Ybf_{i}^{(n)} \|_\mathsf{F}/\| \Zbf_i \|_\mathsf{F}$. In addition, we average the presented results over $100$ independent trials.

In our first experiment utilizing synthetic data, we draw the entries of the nonnegative factor matrices $\Xbf \in \Rbb^{30 \times 5}$ and $\Ybf \in \Rbb^{5 \times 200}$ independently from exponential distributions with parameter values $0.033$ and $0.8$, respectively. We calculate the data matrix as $\Zbf = \Xbf \Ybf + \boldsymbol{\Gamma}$, where we draw the entries of $\boldsymbol{\Gamma}$ independently from a Gaussian distribution with zero mean and variance $3.6 \times 10^{-4}$, resulting in an SNR of approximately $20$dB. We set $\mu = 0.1$, $\eta = 1$, and $g_i = 0.033$ for all agents. We consider $\Zbf$ to be distributed among the agents such that each agent has a varying number of columns between four and $40$. We present the NMSE learning curves of PPDNMF for different values of $N_{\max}$ alongside that of the corresponding centralized algorithm in Fig.~\subref*{fig:1}. We observe from Fig.~\subref*{fig:1} that the proposed PPDNMF algorithm closely approximates the performance of the centralized algorithm in terms of both convergence rate and steady-state NMSE. Additionally, it is also evident that a higher value of $N_{\max}$ leads to a lower steady-state NMSE.

Our second experiment involves the MIT-CBCL face database, which comprises $2429$ monochromic face images in its training set. We distribute the associated data matrix among the agents such that each agent has between $224$ and $245$ columns. For this experiment, we set $\mu = 2$, $\eta = 2$, and $g_i = 0.05$ for all agents. The results presented in Fig.~\subref*{fig:2} underscore the effectiveness of PPDNMF. Notably, PPDNMF exhibits robust performance even with $N_{\max}=10$. Furthermore, we compare the original faces \#1 and \#2429 and their reconstructed versions by PPDNMF using $N_{\max} = 10^6$ in Fig.~\subref*{fig:3}. The reconstructed faces closely resemble their original counterparts.

\section{Conclusion} \label{sec:conc}

We introduced a novel privacy-preserving distributed nonnegative matrix factorization algorithm that employs the Paillier cryptosystem to enable secure collaboration among agents, thereby safeguarding their privacy and mitigating the risk of sensitive data leakage over ad-hoc networks. Our simulation results, based on both synthetic and real data, confirmed the efficacy of the proposed algorithm. In future work, we plan to conduct a comprehensive theoretical privacy analysis of the proposed algorithm, exploring its resilience across various attack scenarios.

\bibliographystyle{IEEEtran}
\bibliography{biblio}

\end{document}